\documentstyle[aps,preprint,prb,epsfig,rotating]{revtex}
\draft
\title{Magnetization of noncircular quantum dots}

\author{Ingibj{\"o}rg Magn{\'u}sd{\'o}ttir\footnote{Present address: 
        Research Center COM, Technical University of Denmark,
        Building 349, DK-2800 Lyngby, Denmark.}  and Vidar Gudmundsson}
\address{Science Institute, University of Iceland, 
         Dunhaga 3, IS-107 Reykjavik, Iceland}
\begin{document}
\maketitle
\tighten
\begin{abstract} 
      We calculate the magnetization of quantum dots deviating
      from circular symmetry for noninteracting electrons or
      electrons interacting according to the Hartree approximation.
      For few electrons the magnetization is found to depend on their   
      number, and the shape of the dot. The magnetization is an
      ideal probe into the many-electron state of a quantum dot.   
\end{abstract}


\section{Introduction}
The internal electron structure of quantum dots has been
explored by far-infrared (FIR) 
absorption,\cite{Demel90:788a}
Raman scattering,\cite{Schueller96:xx} 
and tunneling.\cite{McEuen91:1926,Ashoori92:3088,Blick96:7899} 
The FIR absorption can only be used to detect center-of-mass
oscillations of the whole electron system if it is 
parabolically confined in a circular quantum dot that
is much smaller than the wavelength of the incoming 
radiation, i.e.\ the absorption is according to the extended Kohn 
theorem.\cite{Maksym90:108,Broido90:11400,Peeters90:1486,Chaplik90:31}
Collective excitations due to relative motion of the electrons
are observed in dots where either the circular symmetry 
is broken,\cite{Demel90:788a} 
the radial confinement is not parabolic,\cite{Gudmundsson95:17744} 
or Raman scattering is used to inelastically transfer finite momentum
with a radiation of shorter wavelength to the electron system.
A similar blocking effect has been found in the transport 
excitation spectroscopy
of a quantum dot, where correlation between the multi-electron states
suppresses most of the energetically allowed tunneling processes
involving excited dot states.\cite{Pfannkuche95:1194}
Transitions in Raman scattering measurements do follow certain
selection rules but there is no general mechanism blocking 
a wide selection of collective modes comprized of relative motion
of the electrons. Together these three methods
have proven invaluable for the exploration of excitations in
quantum dots and have supplied important indirect information about the ground
state of the electron system. 

Recently it has been realized that 
new\cite{Grundler98:693,Meinel98a,Meinel99:819}
or improved methods\cite{Wiegers97:3238}
to measure the magnetization of a two-dimensional electrons gas
(2DEG) give important information about the structure of the
many-electron ground state. We are aware of an effort to extend 
these measurements to nanostructured 2DEG's, arrays of quantum wires
or dots. 

The magnetization has been calculated for large ''semi-classical``
circular quantum dots\cite{Fogler94:13767,Tan99:5626}
with noninteracting electrons, or two ''exactly`` interacting
electrons in a square quantum dot with hard walls
and an impurity in the center to investigate how it changes the effect
of the electron-electron interaction on
the magnetization.\cite{Sheng98:152}

In this paper we show that magnetization measurements
of isolated dots can reveal information about the shape of a
dot and the number of electrons in it. In addition, the magnetization
is shown to be very sensitive to the electron-electron interaction.

\section{Model}
We consider a very general angular shape of the quantum dot, thus 
neither the total angular momentum of the system 
nor the angular momentum of the
effective single particle states in a mean field approach is
conserved. The Coulomb interaction 'mixes` up all elements
of the functional basis used and we limit ourselves to the 
Hartree approximation (HA) in order to be able to calculate
the magnetization for up to five electrons.
The confinement potential of the electrons in the quantum dot
is expressed as
\begin{equation}
      V(r,\phi )=\frac{1}{2}m^*\omega_0^2r^2\Big[1+\sum_{p=1}^{p_{max}}
      \alpha_p\cos(2p\phi )\Big], 
\end{equation}
representing an elliptic confinement when $\alpha_1\neq 0$ and
$\alpha_p=0$ for $p\neq 1$, and a square symmetric confinement when
$\alpha_2\neq 0$ and $\alpha_p=0$ for $p\neq 2$. We use the Darwin-Fock
basis; the eigenfunctions of the circular parabolic confinement
potential in which the natural length scale, $a$, is given by
\begin{equation}
      a^2=\frac{\ell ^2}{\sqrt{1+4(\frac{\omega_0}{\omega_c})^2}},
      \quad \ell ^2=\frac{\hbar c}{eB},
\end{equation}
where $\omega_c=eB/m^*c$ is the cyclotron frequency of an electron with
an effective mass $m^*$ in a perpendicular homogeneous magnetic field
$B$. The states are labelled by the radial quantum number
$n_r$ and the angular quantum number $M$.\cite{Gudmundsson91:12098} 

The total magnetization with an orbital contribution $M_o$ defined in 
terms of the quantum thermal average of the current density, and the spin 
contribution $M_s$ derived from the average value of the 
spin density is defined as
\begin{equation} 
      M_o+M_s=\frac{1}{2}\int_{{\bf R}^2}d{\bf r}
      ({\bf r}\times\langle{\bf J(r)}\rangle)\cdot{\bf\hat{n}}
      -g\mu_B\int_{{\bf R}^2}d\bf {r}\langle\sigma_z(\bf {r})\rangle,
\label{Mo+Ms}
\end{equation}
where $\mu_B$ is the Bohr magneton.
The equilibrium local current is evaluated as the quantum
thermal average of the current operator,
\begin{equation}
      \hat{{\bf J}}=-\frac{e}{2}\bigg(\,\hat{{\bf v}}|{\bf r} 
      \rangle\langle{\bf r}| 
      +|{\bf r}\rangle\langle{\bf r}|\hat{{\bf v}}\,\bigg) ,
\label{J_op}
\end{equation}
with the velocity operator 
$\hat{\bf v}=[\hat{{\bf p}}+(e/c)\hat{\bf A}({\bf r})]/m^*$,
$\hat{\bf A}$ being the vector potential. For the finite electron
system of a quantum dot the total magnetization can equivalently
be expressed via the thermodynamic formula 

\begin{equation}
      M_o+M_s=-\frac{\partial }{\partial B}(E_{\mbox{total}}-TS),
\label{Mo+Ms_Th}
\end{equation}
where $S$ and $E_{\mbox{total}}$ are the entropy of the system
and its total energy, respectively. In GaAs $M_s$ is a small
contribution and within the HA it is a trivial 
one.\cite{Gudmundsson99:xxy} Thus, we neglect the spin 
degree of freedom here, but admittedly the spin can be of
paramount importance in connection with exchange effects
on the orbital magnetization.\cite{Gudmundsson99:xxy} 

In order to check the numerical
results we have verified that both definitions (\ref{Mo+Ms}) and
(\ref{Mo+Ms_Th}) give identical results within our numerical  
accuracy for $T=1$ K, even when
the entropy term of (\ref{Mo+Ms_Th}) is neglected.   

\section{Results}
In the numerical calculations we use GaAs parameters, $m^*=0.067m_0$,
and $\kappa =12.4$. Furthermore, we select the confinement 
frequency $\hbar\omega_0=3.37$ meV in order to study quantum dots
with few electrons in the regime where the energy scale of the
Coulomb interaction is of same order of magnitude or larger than
the quantization energy due to the geometry and the magnetic field. 

\begin{figure}[htb]
\begin{center}
      \epsfig{file=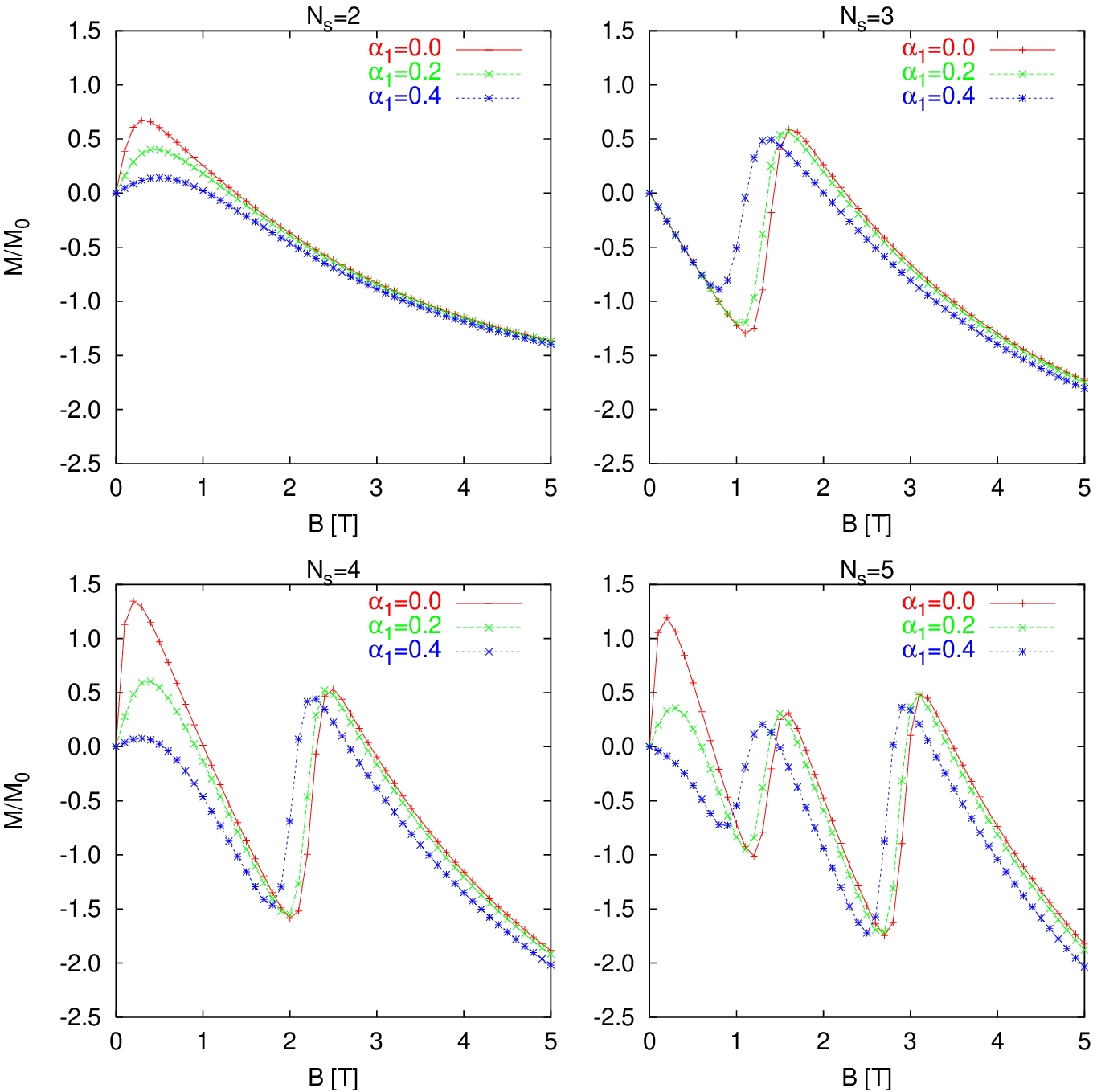,width=15.4cm}
\end{center}
\caption{The effects of increased elliptical deviation 
         on the magnetization of 2-5 
         noninteracting electrons in a quantum dot.
         $M_0=\mu_B=e\hbar/(2m^*c)$, $T=1$ K, $\hbar\omega_0=3.37$ meV. }
\label{fig1}
\end{figure}

\begin{figure}[htb]
\begin{center}
      \epsfig{file=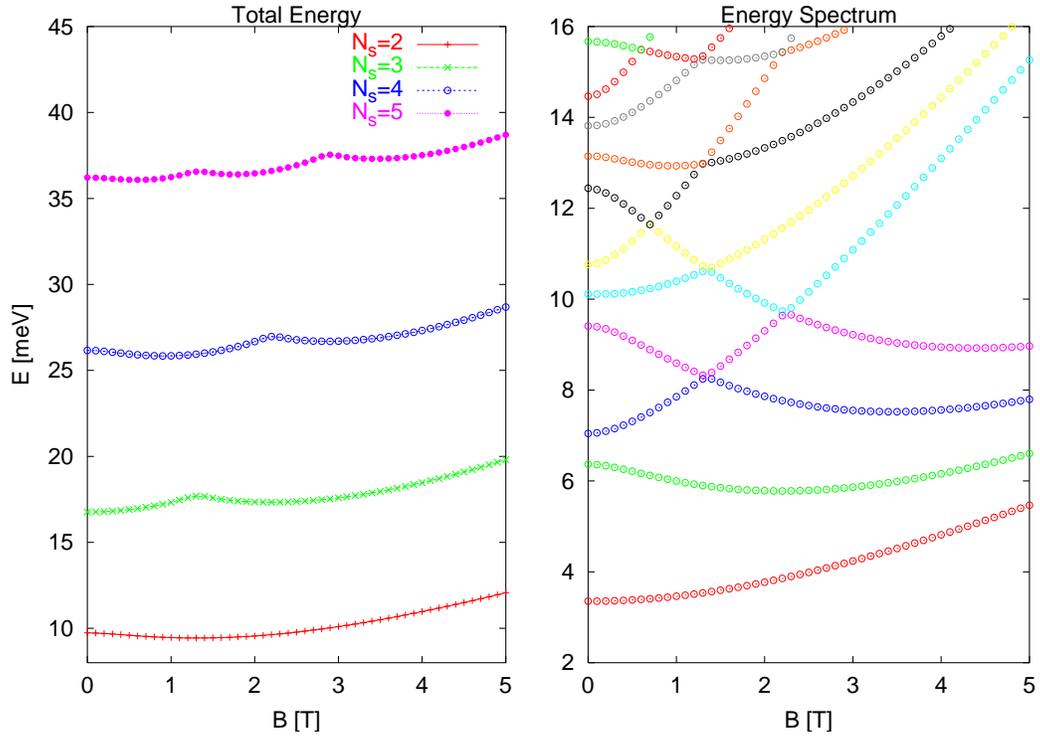,width=14.0cm}
\end{center}
\caption{The total energy (left), and the single-electron energy 
        spectrum (right) for noninteracting electrons in
        an elliptic dot with $\alpha_1=0.2$.}
\label{fig2}
\end{figure}

To make clear the information about the structure of the
ground state discernible in the curves of the magnetization 
versus the magnetic field $B$ we start by investigating
a dot with noninteracting electrons. 
The magnetization of 2-5 electrons in an elliptic quantum dot
is shown in Fig.\ \ref{fig1} for different degree of deviation
from a circular shape. For comparison the total energy 
$E_{\mbox{total}}$ for the same number of electrons, and the
single-electron spectrum for a dot with an elliptical shape
is presented in Fig.\ \ref{fig2}. Sharp jumps in the magnetization
can be correlated with ''discontinuities`` in the derivative 
of $E_{\mbox{total}}(B)$ reflecting crossing of single-electron
states. A jump represents a change in the electron structure 
of the dot. In a circular dot, or a nearly circular dot, each 
single-electron state can be assigned a definite quantum number
$M$ for the angular momentum. As the magnetic field is increased
the occupation of a state with a higher angular momentum is 
energetically favorable.\cite{Pfannkuche93:2244,Maksym90:108}
The mean value of the radial electron density moves away from the
center, the moment of inertia increases, and in order to conserve
the total energy the equilibrium current is reduced leading to
weaker magnetization. In addition to this over-simplified
semiclassical picture the persistent equilibrium current, 
even in a circular dot, has a nontrivial structure as a function 
of the radial coordinate $r$ that can lead to the sign change 
of the magnetization.\cite{Lent91:4179}

For which range of $B$ the effects of the geometry 
are strongest can be understood in the following way:  
For low magnetic field $B\sim 0$ the increased elliptic shape
results in a change of the curvature of some of the single-electron 
energy levels, i.e.\ those that are degenerate for circular dots.
If we look at the magnetization for $N_s=2$ it only differs at
low $B$ for different $\alpha_1$ reflecting the fact that the lowest
occupied single-electron state has almost unchanged curvature 
for low $B$, but the second state is affected. In the case of
three electrons the change in the curvature of the second and the
third state for low $B$ cancels leaving the magnetization 
unaffected by the change in the shape for low magnetic field.
Instead, the magnetization changes with $\alpha_1$ around
$B\sim 1$ T where the third and the fourth single electron state
cross. The crossing point varies with $\alpha_1$ shifting the
location of the jump in the magnetization. In this same sense
all the variation in the magnetization can be referred back to 
the single electron energy spectrum, and thus the total energy.

\begin{figure}[htb]
\begin{center}
      \epsfig{file=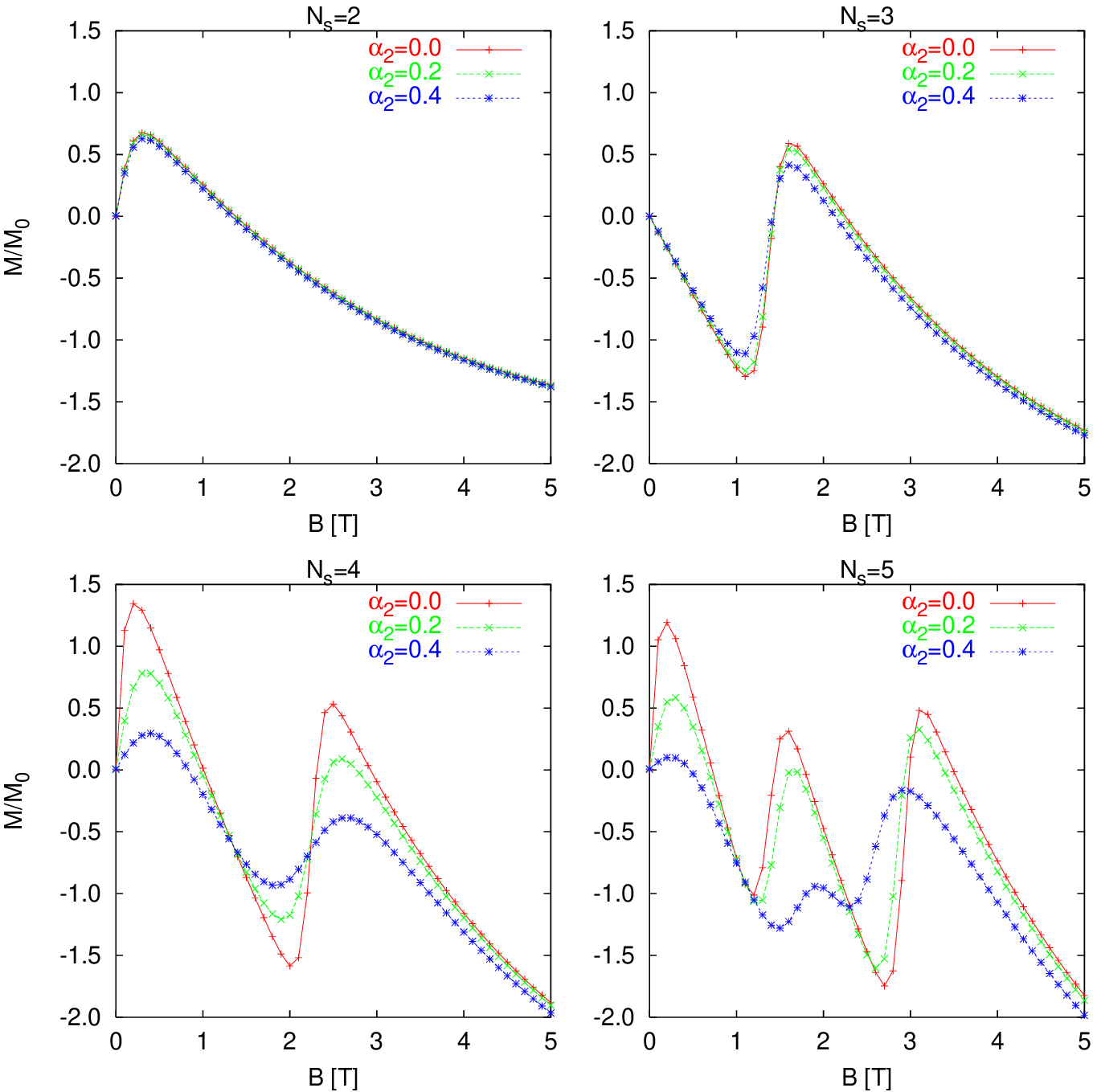,width=15.4cm}
\end{center}
\caption{The effects of increased square deviation 
         on the magnetization of 2-5 
         noninteracting electrons in a quantum dot.
	$M_0=\mu_B=e\hbar/(2m^*c)$, $T=1$ K, $\hbar\omega_0=3.37$ meV. }
\label{fig3}
\end{figure}

\begin{figure}[htb]
\begin{center}
      \epsfig{file=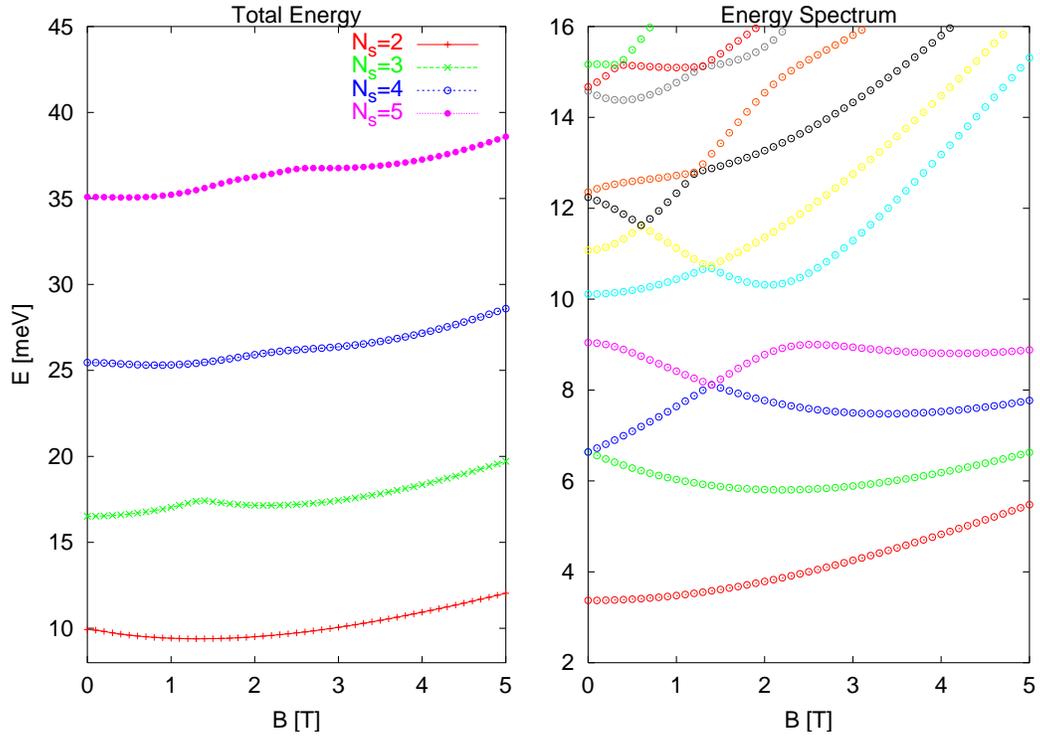,width=14.0cm}
\end{center}
\caption{The total energy (left), and the single-electron energy 
        spectrum (right) for noninteracting electrons in
        a square dot with $\alpha_1=0.4$.}
\label{fig4}
\end{figure}

So, what changes do we observe when we change the degree of
a square deviation of a quantum dot instead of the elliptical
shape? The corresponding graphs for the magnetization and the
energy can be seen in Fig.\ \ref{fig3} and \ref{fig4}, respectively
in case of square deviated dots.
The square deviation does not lift the degeneracy of the 
second and the third single-energy levels at $B=0$ and it does
not move strongly the crossing point between the third and the
fourth energy level. The magnetization for $N_s=2$ and $3$ 
is thus not strongly effected by increased square shape of a
quantum dot. On the other hand, the magnetization is strongly 
affected by the change in the shape of dots with four or
five electrons. The large change in the anticrossing seen in
the single-energy spectrum (at $B\approx 2.3$ T) with increasing
$\alpha_2$ is clear in the magnetization. 

The main difference in the magnetization of a quantum dot
with an elliptical or square shape comes from the fact that
the elliptical deviation has nonzero matrix elements between
single electron states with a dominant contribution of
basis states of a circular dot satisfying $|\Delta M|=2$,
whereas the square shape can only connect states with
the dominant contribution satisfying $|\Delta M|=4$. For
weak deviation the square shaped dot thus needs the occupation of
more states than the elliptical one to show effects in the 
magnetization different from the magnetization of a circular
quantum dot.  

\begin{figure}[htb]
\begin{center}
      \epsfig{file=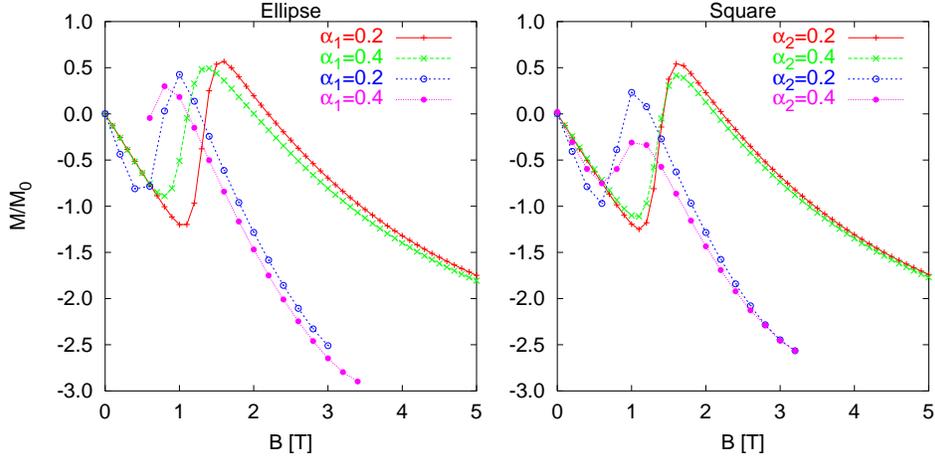,width=15.4cm}
\end{center}
\caption{The magnetization of a three electron quantum dot in the case 
        of elliptic  confinement (left), with $\alpha_1=0.2$ and $0.4$, 
        and square symmetric confinement (right), with $\alpha_2=0.2$, 
        and $0.4$ for both interacting (closed symbols), 
        and non-interacting electrons (open symbols).
	$M_0=\mu_B=e\hbar/(2m^*c)$, $T=1$ K, $\hbar\omega_0=3.37$ meV.}
\label{fig5}
\end{figure}

\begin{figure}[htb]
\begin{center}
      \epsfig{file=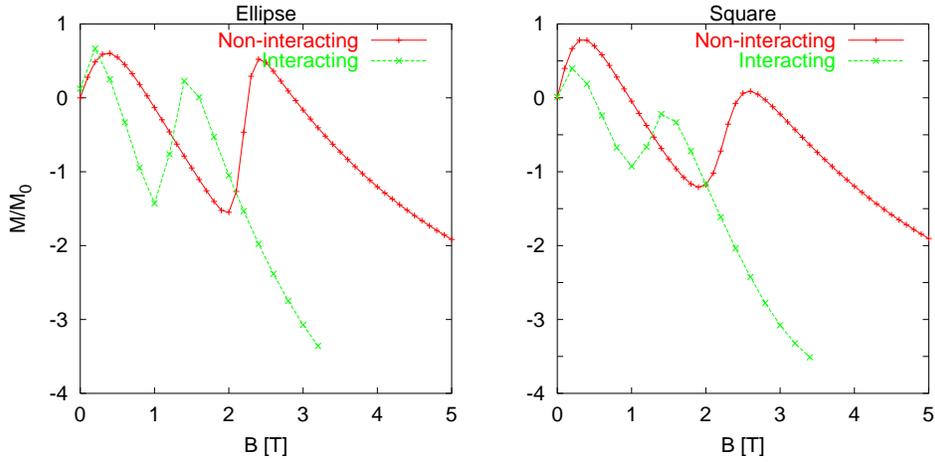,width=15.4cm}
\end{center}
\caption{The magnetization of a four electron quantum dot in the case 
        of elliptic  confinement (left), with $\alpha_1=0.2$, 
        and square symmetric confinement (right), with $\alpha_2=0.2$, 
        for both interacting and non-interacting electrons.
	$M_0=\mu_B=e\hbar/(2m^*c)$, $T=1$ K, $\hbar\omega_0=3.37$ meV.} 
\label{fig6}
\end{figure}

Within the Hartree approximation the essential structure of the
effective single electron spectrum remains, i.e.\ the anticrossing
typical for the square shape and the lifting of the degeneracy of
the second and third energy level typical for the elliptic shape.
The finer details of the spectrum depend on the number of 
electrons in the dot as is expected in an effective potential.
The electron-electron interaction in a circular dot 
can not change the angular symmetry of the dot, but as soon as
this symmetry is broken by the confinement potential the 
interaction further modifies the angular shape of the dot. 
In Fig.\ \ref{fig5} and \ref{fig6} the magnetization for three
and four interacting electrons is compared to the magnetization
of noninteracting electrons. The Coulomb repulsion between the
electrons causes changes in the electron structure to occur
for lower magnetic field $B$. It is energetically favorable for 
electrons to occupy states associated with higher angular momentum
earlier when $B$ is increased in order to reduce the overlap of their
wavefunctions. The jumps in the magnetization are thus shifted 
toward lower $B$. 

In the noninteracting case the magnetization of three
electrons in a square shaped dot did not vary much with increased
deviation from a circular shape, see Fig.\ \ref{fig3}. Quite
the opposite happens for three interacting electrons in the
same system, the reason being that the interaction enhances 
the deviation and thus the size of the anticrossing energy gap 
in the effective single electron spectrum. This behavior is only 
observed in dots with few electrons, in larger dots the interaction
generally smoothens the angular shape of the dots.

\section{Summary}
The calculation of the magnetization for few electrons in 
a quantum dot shows that it depends on the shape of the dot
and the exact number of electrons present. This is in contrast 
to the results for large dots with many electrons where the
magnetization assumes properties reminiscent of a homogeneous
2DEG.\cite{Fogler94:13767} It is certainly harder to justify 
the use of the Hartree approximation here than in the case
of a calculation of the far-infrared absorption where 
to a large degree, classical modes -  center-of-mass 
modes - dominate the excitation spectrum. We are fully aware that
exchange and correlation effects will be important in the 
present system,\cite{Pfannkuche93:2244,Pfannkuche94:1221}
but they will not qualitatively change our results that
magnetization measurements are ideal to investigate the
many-electron structure of quantum dots.

%
%
\acknowledgments
This research was supported by the Icelandic Natural Science Foundation
and the University of Iceland Research Fund. 
In particular we acknowledge additional support from 
the Student Research Fund (I.\ M.). 
%
%
\bibliographystyle{prsty}

%
%
%
%
%
%
%
%
%
%
%
\newpage
%
%
%
%
\end{document}